\documentstyle[12pt]{article}
\title{Detecting New Physics from CP-violating
phase measurements in $B$ decays}
\author{Jo\~ao P.\ Silva$^a$ and L.\ Wolfenstein$^b$ \\
\\
\small $^a$ Centro de F\'{\i}sica Nuclear da Univ. de Lisboa,
Av.\ Prof.\ Gama Pinto, 2\\
\small 1699 Lisboa Codex, Portugal\\
\small and Centro de F\'{\i}sica, ISEL\\
\\
\small $^b$ Department of Physics, Carnegie-Mellon University, \\
\small Pittsburgh, Pennsylvania 15213, U.S.A.}
\begin{document}
\maketitle
\begin{abstract}
The standard CKM model can be tested and New Physics detected using
only CP-violating phase measurements in $B$ decays.
This requires the measurement of a phase factor which is small in
the Standard Model, in addition to the usual large phases
$\beta$ and $\gamma$.
We also point out that identifying violations of the unitarity
of the CKM matrix is rather difficult, and cannot be done with
phase measurements alone.
\end{abstract}
 
 
The major goal of future experiments on B mesons is to measure CP
violation \cite{sanda} to test the Standard Model (SM) based on
the CKM matrix \cite{CKM}, and to detect possible evidence for
physics beyond the SM.
Several articles have discussed this issue focusing on tests of
the general assumptions behind the SM predictions \cite{Nir90a,Nir90b}.
More specific predictions can be made by concentrating on
particular examples of physics beyond the SM \cite{Dib91,Gro96}.
Here we discuss what can be learned exclusively from precision measurements
of CP-violating phases in B decays (without making use of other 
quantitative information), if the new effects are not dominant.

In a standard notation, the CKM matrix can be expanded in powers of 
the Cabibbo angle $\lambda = \sin \theta_c$, as \cite{Wol83},
\begin{equation}
V =
\left( \begin{array}{ccc}
1-\frac{1}{2} \lambda^2 &
\lambda &
A \lambda^3 (\rho - i \eta (1-\frac{1}{2} \lambda^2))\\
- \lambda &
1-\frac{1}{2} \lambda^2 - i \eta A^2 \lambda^4 &
A \lambda^2 (1+i \eta \lambda^2)\\
A \lambda^3 (1- \rho - i \eta) &
- A \lambda^2 &
1
\end{array} \right)\ ,
\label{eq:ckmmatrix}
\end{equation}
with the expansion truncated when unitarity is satisfied to order $\lambda^3$
for the real part and to order $\lambda^5$ for the imaginary part.
We will assume the general hierarchical structure of 
Eq.~(\ref{eq:ckmmatrix}),
but there are no assumptions about magnitudes other than that $\rho$,
$\eta$, and $A$ are less than unity.
In fact, some of our analysis does not even depend
on the approximate magnitudes of the $V_{ub}$, $V_{td}$ and $V_{ts}$ 
matrix elements, which are poorly measured.

It is pointed out in an article by Aleksan, Kayser and London
\cite{Ale94} that the matrix contains only four independent
phases, which may in principle be determined from CP violating
experiments. While they emphasize the possibility of reconstructing
the matrix from these four phases, our goal is to use them to detect
new physics. In the SM, only two of these phases are large;
these are essentially the ones usually identified as $\beta$ and $\gamma$.
With our weak constraints on $\rho$ and $\eta$,
the angles $\beta$ and $\gamma$
can have almost any value. Therefore, these measurements by themselves
provide practically no test of the CKM model.
We recall that, in the standard analysis, one combines the measurements of 
$|V_{ub}|$, $|V_{td}|$ and the CP violation in the neutral Kaon system
to constrain the allowed values of $\beta$ and $\gamma$ \cite{Ali96}.
New physics would then show up through novel correlations between 
different experiments \cite{Soa93,Nir93}.
Our major emphasis is on what can be learned by attempts to measure a
third phase $\epsilon$ (not to be confused with the parameter in
$K$ decays) which is expected to be much smaller \cite{Ale94}.

We follow reference \cite{Ale94} and define the two large phases as
\begin{eqnarray}
\beta & = &
\arg \left( - \frac{V_{tb} V_{td}^\ast}{V_{cb} V_{cd}^\ast}
\right)
\label{eq:beta}\ ,
\\
\gamma & = &
\arg \left( - \frac{V_{ub}^\ast V_{ud}}{V_{cb}^\ast V_{cd}}
\right)
\label{eq:gamma}\ .
\end{eqnarray}
Within the SM, any other large phase we might chose will differ
from these only by a term of order $\epsilon$, defined as
\begin{equation}
\epsilon  = 
\arg \left( - \frac{V_{cs}^\ast V_{cb}}{V_{ts}^\ast V_{tb}}
\right)
\label{eq:epsilon}\ .
\end{equation}
The last phase needed,
\begin{equation}
\epsilon^\prime  = 
\arg \left( - \frac{V_{ud}^\ast V_{us}}{V_{cd}^\ast V_{cs}}
\right)
\label{eq:epsilon'}\ ,
\end{equation}
is much smaller than the others, in the SM.

Aleksan, Kayser and London now make the important point that,
to a good approximation, we can check the CKM model from the
equation \cite{Ale94}
\begin{eqnarray}
\sin{\epsilon} & \simeq &
{\left| \frac{V_{us}}{V_{ud}} \right|}^2
\frac{\sin{\beta}\  \sin{\gamma}}{\sin{(\beta + \gamma)}}
\nonumber\\
   & \simeq &
\lambda^2 \eta
\label{eq:master}\ ,
\end{eqnarray}
where the last equality follows from Eq.~(\ref{eq:ckmmatrix}).
The approximation involves corrections which, percentage-wise,
are at most of order $\lambda^2$. 
The power of this relation lies in the fact that the ratio
$|V_{us}/V_{ud}|$ is known to high precision.
Similar relations may be derived using other magnitude ratios,
like $|V_{cd}/V_{cs}|$, which are not so well determined.
There are two other sets of expressions
involving only $\beta$, $\gamma$, and $\epsilon$, but they require
the knowledge of $|V_{ub}|$ or $|V_{td}|$.
We note that the validity of Eq.~(\ref{eq:master}) rests on two pillars.
On the one hand, it assumes that the extraction of the angles was not
inhibited by new physics effects in $B_d - \bar{B}_d$,
or $B_s - \bar{B}_s$ mixing.
On the other hand,
it explicitly uses unitarity when relating the angles with the
magnitudes of CKM matrix elements.

We now turn to the use of Eq.~(\ref{eq:master}) to detect
physics beyond the Standard Model.
In Tables \ref{tab:Bd} and \ref{tab:Bs}
%
%
%
%
%
%
\begin{table}[hbt]
\centering
\begin{tabular}{||c||c|c|c||}
\hline 
\hfil Class\hfil& \hfil sub-process\hfil &
\hfil channel\hfil &\hfil CP asymmetry\hfil 
\\\hline\hline
1d & $\bar{b} \rightarrow \bar{c}c\bar{s} $& 
$\psi K_s$ & $- \sin{(2 \beta - \theta_d)}$ 
\\
2d & $\bar{b} \rightarrow \bar{c}c\bar{d} $& 
$D^+ D^-$ & $- \sin{(2 \beta - \theta_d)}$ 
\\
3d & $\bar{b} \rightarrow \bar{u}u\bar{d} $& 
$\pi^+ \pi^-$ & $- \sin{(2 \beta +2 \gamma - \theta_d)}$ 
\\
4d & $\bar{b} \rightarrow \bar{s}s\bar{s} $& 
$\phi K_s$ & $- \sin{(2 \beta +2 \epsilon - \theta_d)}$ 
\\ \hline 
\end{tabular}
\caption{CP violating asymmetries in $B_d$ decays.}
\label{tab:Bd}
\end{table}
%
%
%
%
%
%
\begin{table}[hbt]
\centering
\begin{tabular}{||c||c|c|c||}
\hline 
\hfil Class\hfil& \hfil sub-process\hfil &
\hfil channel\hfil &\hfil CP asymmetry\hfil 
\\\hline\hline
1s & $\bar{b} \rightarrow \bar{c}c\bar{s}$ & 
$D_s^+ D_s^-$ & $ \sin{(2 \epsilon + \theta_s)}$ 
\\
2s & $\bar{b} \rightarrow \bar{c}c\bar{d}$ & 
$\psi K_s$ & $ \sin{(2 \epsilon + \theta_s)}$ 
\\
3s & $\bar{b} \rightarrow \bar{u}u\bar{d}$ & 
$\rho K_s$ & $- \sin{(2 \gamma - 2 \epsilon - \theta_s)}$ 
\\
4s & $\bar{b} \rightarrow \bar{s}s\bar{s}$ & 
$\eta^\prime \eta^\prime$ & $\sin{(\theta_s)}$ 
\\ \hline 
\end{tabular}
\caption{CP violating asymmetries in $B_s$ decays.}
\label{tab:Bs}
\end{table}
we list a set of CP violation experiments and indicate what is measured
in terms of $\beta$, $\gamma$, and $\epsilon$.
In these tables we have allowed
for the most probable type of new physics, namely, that which
contributes new phases to $B$ - $\bar{B}$ mixing.
This will add $\theta_d$ to the SM $B_d - \bar{B}_d$ mixing phase,
and $\theta_s$ to the SM $B_s - \bar{B}_s$ mixing phase.
Unitarity is not assumed in calculating these CP asymmetries,
but it is assumed that the
decays are dominated by intermediate $W$ bosons, and that there
are no detectable new phases in the $K$ system.
We have classified the decays as in reference ~\cite{Nir90b}.
Decays based on the quark sub-process
$\bar{b} \rightarrow \bar{u}u\bar{s}$ have not been included for
they have similar contributions from tree and penguin diagrams.
Similarly, those decays involving the quark sub-process
$\bar{b} \rightarrow \bar{s}s\bar{d}$ were dropped,
since they are likely to be more affected by penguin diagrams with
virtual charm and up quarks.
Note that the asymmetries 1 and 2 are equal \cite{Nir90a}.
This is due to the $K - \bar{K}$ mixing phase which is mandatory
in order for the result to be invariant under a rephasing of the
$s$ and $d$ quarks.

The first channels to be measured at $B$-factories are 
$\psi K_s$ and\footnote{
It is well known that this last measurement is obscured by the
presence of a small penguin contribution
\cite{penguins}.
This can be overcome measuring isospin related channels
\cite{Gro90}, though that requires the experimentally challenging
detection of $\pi^0$'s. Using the $K \pi$, SU(3) related channels
\cite{Sil94} will be easier, since most proposed detectors have 
good charged meson identification.
}
$\pi^+ \pi^-$.
This will permit a correct determination of $2 \gamma$, but $2 \beta$ 
appears always in connection with $\theta_d$. 
The first measure of $\epsilon$ is likely to come from process (2s),
$B_s \rightarrow \psi K_s$. 
A failure of Eq.~(\ref{eq:master}) would then mean that
$\theta_s \neq 0$, and we did not correctly measure $\epsilon$,
or that $\theta_d \neq 0$, and we have the wrong value for $\beta$,
or that the values of $\epsilon$ and $\beta$ are correct, but 
the CKM matrix is not $3 \times 3$ unitary, so that the sides and angles
are not related in the expected way.
It could be a combination of all three.
Of course,
as was stressed by Nir and Silverman \cite{Nir90a},
if the asymmetry of process (2s) is much larger than $\lambda^2$,
there must be a new contribution $\theta_s$ to the $B_s$ - $\bar{B}_s$
mixing phase.
Our analysis is directed to smaller violations of Eq.~(\ref{eq:master}).

For the moment we concentrate on the mixing effects assuming that
the nonunitarity is less important. We will come back to it later on.
To distinguish the cases of $\theta_s \neq 0$ from $\theta_d \neq 0$,
it is necessary to learn about the phases of $V_{ts}$ or $V_{td}$
from sources other than mixing.
The most likely processes are the
decays of the form $\bar{b} \rightarrow \bar{s}s\bar{s}$,
in classes (4d) and (4s).
These are expected to be dominated by a penguin graph proportional
to $V_{tb} V_{ts}^\ast$. Note that the penguin graphs involving 
$u$ and $c$ quarks are suppressed by an extra power of $\lambda^2$.
The asymmetry of (4d) compared to (1d) gives the true value of
$\epsilon$. This yields two pieces of information.
We can insert this correct value of $\epsilon$ into Eq.~(\ref{eq:master}),
so that a failure of the equality must then be due to $\theta_d$.
Moreover, if this value of $\epsilon$ differs from that deduced from (2s),
then there is a non-zero value of $\theta_s$, which could also be directly
detected from the asymmetry of (4s).

If the new physics is superweak \cite{Wol64}, it might make a
similar contribution to $B_d - \bar{B}_d$ and $B_s - \bar{B}_s$
mixing. Suppose that contribution is of the order
of $\Delta M (B_d)$. The result is a large value for $\theta_d$,
but a value of order $\lambda^2$ for $\theta_s$.
To detect such a value for $\theta_s$ requires determining
$\epsilon$ to an accuracy of a fraction of $\lambda^2$.
From the present analysis, a large value of $\theta_d$ has the effect
of giving the wrong value of $\beta$ to insert in Eq.~(\ref{eq:master}),
and thus changes the calculated value of $\epsilon$ by a term of
order $\lambda^2$. This again requires determining $\epsilon$ to high
accuracy.
Thus, in the absence of quantitative knowledge of the magnitudes of
$V_{ub}$ and $V_{td}$, the use of CP-violating phases alone to
detect new physics is likely to require very precise measurements.

While these arguments hold if the new physics is superweak
\cite{Wol64}, in many theories there may be significant new contributions
to $\bar{b} \rightarrow \bar{s}s\bar{s}$. The importance of such new
penguin-type diagrams has been emphasized in discussions of the decay
$b \rightarrow s \gamma$ \cite{hewett},
and, more recently,
in the comprehensive study of Gronau and London \cite{Gro96}.

We now turn to the question of identifying violations of the unitarity
of the $3 \times 3$ CKM matrix,
as can occur, for example, in models with extra quarks.
It is often suggested that one can test unitarity by measuring
three large phases $\tilde{\beta}$, $\tilde{\alpha}$, and 
$\tilde{\gamma}$ from reactions (1d), (3d), and (3s), respectively,
and see if they add up to $\pi$.
Following reference \cite{Ale94},
we have emphasized that there are only two large angles,
$\beta$ and $\gamma$,
and that the third relevant angle is the small angle 
$\epsilon$.
Unitarity then implies Eq.~(\ref{eq:master}).
However,
the failure of Eq.~(\ref{eq:master}) can be attributed to the presence
of $\theta_d$, as discussed above, and thus does not imply a failure of
unitarity.
Indeed a large class of theories,
including those in which unitarity is violated,
give a significant value of $\theta_d$.
We thus reach the conclusion that it is impossible to identify
a violation of unitarity simply from measurements of three independent
CP-violation phases.

We stress that deviations from unitarity will show up in a
failure of SM relations involving {\it both} angles and magnitudes.
In particular, the relation
$\tilde{\alpha} + \tilde{\beta} + \tilde{\gamma} = \pi$ does
{\it not} test unitarity.
Such a test will only arise when one confronts these angles with
the relevant magnitudes in the unitarity triangle.

Moreover,
clearly identifying unitarity violations,
even in principle,
requires precise knowledge of CKM magnitudes other than
the Cabibbo angle.
It is then possible to derive various relations between angles
and magnitudes independent of Eq.~(\ref{eq:master}).
However, unless both $|V_{td}|$ and $|V_{ub}|$ are included,
it is still necessary to have a measurement of the small phase
$\epsilon$.
An example of such a relation is 
\begin{equation}
\sin{\epsilon} \simeq
\left\{
1 - \left| \frac{V_{cs} V_{cb}}{V_{ts} V_{tb}} \right|
\right\}\ 
\tan{\gamma}\ .
\label{eq:do1}
\end{equation}
This would distinguish the two effects.
There is a similar relation involving the same magnitudes and $\beta$,
but that would not solve the problem of disentangling
$\theta_d$ from nonunitarity.
Note that the order to which we have taken the Wolfenstein parameterization
of Eq.~(\ref{eq:ckmmatrix}), is not good enough to confirm this relation.
This is due to the fact that we need to know 
$|V_{cs} V_{cb}|/|V_{ts} V_{tb}|$ to order $\lambda^2$,
requiring very precise measurements.

In conclusion,
we have shown that the detection of new physics exclusively
through CP asymmetries in $B$ decays,
requires the measurement of $2 \beta - \theta_d$,
$\gamma$ and a third angle $2 \epsilon + \theta_s$.
Unless the new phase in $B_s - \bar{B}_s$ is large \cite{Nir90a},
such analysis requires a precision down to $\epsilon \simeq \eta \lambda^2$.
A failure of Eq.~(\ref{eq:master}) will then signal new physics.
Whether this is due to $\theta_s$, will be determined once an
asymmetry measurement is made in a
$\bar{b} \rightarrow \bar{s}s\bar{s}$ decay.
However, disentangling $\theta_d$ from unitarity violations requires 
precision measurements of CKM magnitudes, other than the Cabibbo angle.

 
\vspace{5mm}
 
J.\ P.\ S.\ is indebted to Carnegie Mellon University's Department of Physics
for their kind hospitality while this work was done.
The work of L.\ W.\ was supported by the United States Department of Energy,
under the contract DE-FG02-91ER-40682. The work of J.\ P.\ Silva was
supported by the Portuguese JNICT under project CERN/P/FAE/1050/95.


%

\end{document}